# A Moving Magnetic Trap Decelerator: a New Source for Cold Atoms and Molecules

Etay Lavert-Ofir, Sasha Gersten, Alon B. Henson, Itamar Shani, Liron David, Julia Narevicius and Edvardas Narevicius

Department of Chemical Physics, Weizmann Institute of Science, Rehovot 76100, Israel

**We present an experimental realization of a moving magnetic trap decelerator, where paramagnetic particles entrained in a cold supersonic beam are decelerated in a co-moving magnetic trap. Our method allows for an efficient slowing of both paramagnetic atoms and molecules to near stopping velocities. We show that under realistic conditions we will be able to trap and decelerate a large fraction of the initial supersonic beam. We present our first results on deceleration in a moving magnetic trap by bringing metastable neon atoms to near rest. Our estimated phase space volume occupied by decelerated particles at final velocity of 50 m/s shows an improvement of two orders of magnitude as compared to currently available deceleration techniques.**

Cold chemistry (at temperature of ~1K) is a fast developing field with many interesting theoretical predictions and only a few experiments[1-4]. It has been predicted that in the cold chemistry regime reactions might proceed through resonance states[5] and both reaction rate and path can be controlled[6] via external magnetic or electrostatic fields. Low temperature chemistry is not limited only to artificially created conditions but occur naturally at the low temperature of interstellar space[7] and atmospheres of remote planets and moons[8]. It is anticipated that cold chemistry will follow the enormous progress in a molecular chemistry that was spurred by the advent of supersonic beams more than half a century ago[9]. However, the progress in the experimental cold chemistry is rather slow partly due to the absence of a general method of atomic and molecular cooling. Although low temperature environment created within a supersonic expansion has been used to observe many reactions[10], a direct measurement in a classical crossed beam arrangement has been limited to temperatures of about 20 K[4]. We have developed a new method of supersonic beam deceleration that is applicable to paramagnetic species, both atoms and many molecular radicals. Our method is the first implementation of a moving magnetic trap decelerator that enables slowing of paramagnetic species to stopping velocities with higher efficiencies and numbers. Since our method is easily applicable to paramagnetic atoms we expect that the moving magnetic trap decelerator will make a considerable impact on cold chemistry allowing study of many atom-radical reactions. For example molecular radical reactions with atomic oxygen, nitrogen, hydrogen are central both in astrophysics[7] and combustion chemistry[11].

Many groups realized the benefits of supersonic expansions as a source of cold atoms or molecules and a decade ago the first experiments began exploring the possibility of controlling the mean velocity of supersonic beams. Different approaches have been tested including kinetic methods of control (mechanical [12] or collisional [13]), and methods based on interaction with pulsed electric [14] (applicable to polar molecules with linear Stark effect), magnetic [15, 16] (applicable to paramagnetic atoms and molecules) and laser fields [17]. Another method that is also able to cool paramagnetic atoms and molecules is the buffer gas cooling [18]. Since high densities of inert buffer gas are needed to thermalize the target atoms or molecules the buffer gas cooling method works the best with atoms and molecules with zero orbital angular momentum.

Efficient deceleration of supersonic beams has been a long-standing goal for beam deceleration methods since barrierless reaction rate constants (usually in the range of $10^{-10}$-$10^{-13}$cm$^3$ molecule$^{-1}$ s$^{-1}$) place a stringent requirement on reactants density in the reaction volume. The problem has been extensively addressed in the Stark effect based deceleration methods. During the deceleration polar molecules are confined in a dynamical trap. One must make sure that the longitudinal and transverse degrees of freedom remain decoupled throughout the deceleration process otherwise losses from within the dynamical trap occur. It has been both suggested [19] and experimentally confirmed that Stark decelerators with dynamical confinement can be operated in a high order mode that decouples the longitudinal and transverse motion down to velocities of 150 m/s [20]. At lower velocities the decoupling breaks down and the losses occur from within the dynamical trap. As a result the phase space volume occupied by decelerating polar molecules (acceptance) at 50 m/s approaches the acceptance of the conventional Stark decelerator which is an order of magnitude lower.

Since we are interested in the cold chemistry with both atoms and molecules we focus our attention on the Zeeman effect based deceleration performance at low velocities. It has been suggested that the loss problems at low velocities can be overcome by decelerating in a co-moving three dimensional magnetic trap [21]. Similar approach has been realized with moving electrostatic traps where a beam of metastable CO molecules has been decelerated in a three dimensional trap formed on a microstructured chip [22, 23] and recently a deceleration of metastable CO molecules to 140 m/s final velocity has been demonstrated in a moving macroscopic electrostatic trap [24]. Also Rydberg atoms have been decelerated in a moving electrostatic trap [25].

In this article we present an experimental realization of a moving magnetic trap decelerator idea that was proposed earlier [21]. We show deceleration of metastable Neon that has a mass to magnetic moment ratio similar to atomic oxygen or molecular radicals such as OH and NH from the initial velocity of 430 m/s to final velocity of 54 m/s removing more than 98 % of initial kinetic energy. We also present a numerical analysis of the deceleration process that shows an improvement in the phase space volume (acceptance) occupied by the decelerating particles by two orders of magnitude as compared to state of the art Stark decelerator. Finally we show under which conditions our decelerator will operate with maximal acceptance and decelerate a beam of paramagnetic particles to stopping velocities without loss other than by selection of low field seeking states that are magnetically trappable.

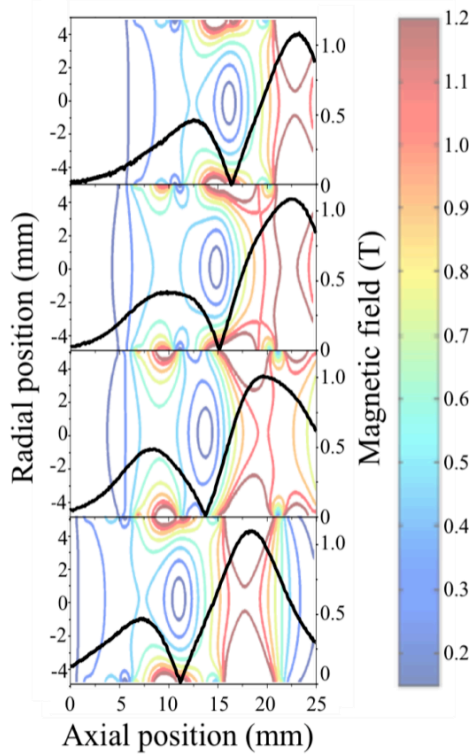

**Figure 1**. Black solid lines are the snapshots of the axial magnetic field as measured during the magnetic trap movement. Trap center moves by 5.2 mm in time steps of 4 μs. The contour plots represent the magnetic fields as calculated by finite element simulation.

Recently, Trimeche et. al demonstrated trapping of a supersonic beam in a travelling magnetic wave[26].

The moving magnetic trap deceleration idea has been inspired by the atom conveyor that has been successfully used to transport atoms on micro chips [27] or across centimeter long distances in cold atom experiments [28]. Although conceptually similar, in our case we need to translate the magnetic trap with far higher velocities. The initial velocity of the moving magnetic trap has to be in the range of a few hundred m/s to match the initial velocity of a supersonic beam. We also require much higher deceleration values. In the case of cold atom transport the acceleration stays in the range of 1 m/s$^2$. In order to decelerate a supersonic beam from 500 m/s to a full stop in 1 meter a deceleration in excess of 100,000 m/s$^2$ is necessary.

We create an effective moving magnetic quadrupole trap potential by constructing a series of spatially overlapping traps and activating them in a temporally overlapping pulse sequence. Each trap consists of two coils in an anti-Helmholtz configuration and is axially shifted from the next trap by half the coil-to-coil distance. Once the magnetic field in a given trap reaches the maximal value we gradually switch the adjacent trap on and shift the magnetic field minimum to its new position in the center of the next trap. The center-to-center distance between the coils forming a single trap is 10.4 mm with a bore diameter of 10.2 mm. Our configuration is not symmetric; one coil has 16 windings whereas the second coil has 8, both of a 0.46 mm diameter copper wire. Out of 213 traps, 161 last trap coils are encased in Permendur shells. With a pulsed current peaking at 500 A the front barrier height exceeds 1 T whereas the back barrier height is 0.4 T (in the lab frame). The magnetic field is lower by 20% for the first 52 traps constructed from coils that have no Permendur shell.

In Fig. 1 we present measurements of the trap translation (solid lines). We measure the time and position dependence of the magnetic field using the Faraday effect [29]. We insert a 1mm thick terbium gallium garnet crystal, mounted on a translation stage, into the bore of our coils and monitor the rotation of a linearly polarized 532 nm laser beam (spot size 100 um) as we pulse currents through our coils.

The overlap time between two consecutive current pulses defines the velocity of the trap at different slowing stages. We change the trap velocity by gradually increasing the overlap time. It includes increasing both the pulse duration and pulse-to-pulse time difference. Each current pulse is formed by an electronic driver which is based on a LCR circuit. In order to control the pulse length, we have designed a 'binary tune box' (BTB) of inductors, which allows us to set the LCR inductance with an accuracy of 1μH. The BTB is composed of 6 inductors ($2^0$, $2^1$... $2^5$μH) connected in series and power transistors which digitally determine through which of the inductors the current of each quadrupole driver flows. As one can see from Fig. 1 the trap center movement is not smooth and the acceleration changes during the trap movement. At a constant velocity of 430 m/s particles experience two opposite sign momentum kicks with a frequency of ~80 kHz. Since the perturbation frequency is almost two orders of magnitude larger than the characteristic frequency of particle motion and the trapping potential is linear with coordinate (quadrupole trap) we can safely treat the motion of a particle in a moving trap using a time averaged potential only [30].

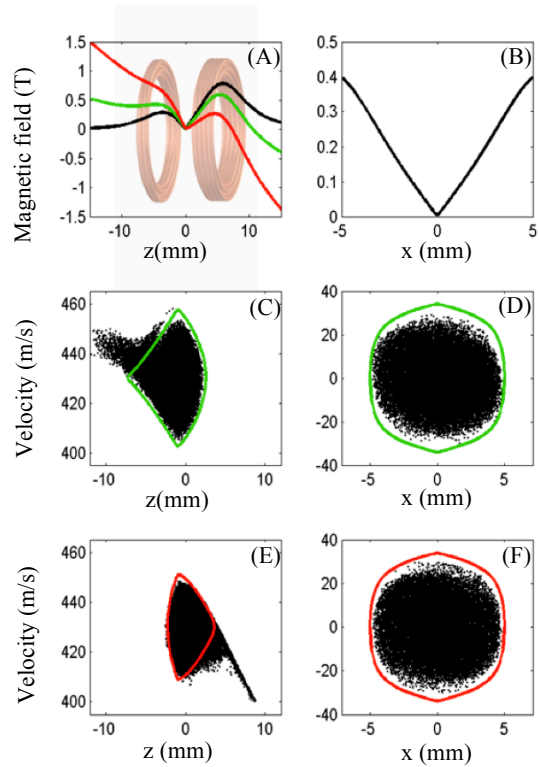

**Figure 2.** (A) solid black line: the time averaged axial magnetic field of a quadrupole trap as calculated by finite element method using experimental parameters (coils without Permendur shell). Solid green (red) line: the average effective magnetic potential that includes tilt due to fictitious force, deceleration from 430 to 350 m/s (50 m/s). (B) The time averaged transverse magnetic field of a quadrupole trap. (C-F) Longitudinal and transverse phase space distributions along with one dimensional separatrices for 350 m/s (C and D) and 50 m/s (E and F) final velocities.

It is convenient to consider the one dimensional approximation to the deceleration process by moving into the decelerating frame of reference. The fictitious force modifies the effective potential and adds a term that is linear in the propagation coordinate, reducing the height of the "front" barrier, as seen in Fig. 2 (A) where we present a canonical magnetic quadrupole trap in both lab and accelerating frames of reference. The initial magnetic field gradient of 160 T/m that we are able to achieve allows us to apply higher decelerations without losing many trapped particles. For example in the case of metastable neon, that has a similar mass to magnetic moment ratio as paramagnetic molecular radicals such as NH or OH in rovibrational ground state, a deceleration of 83,000 m/s$^2$ adds a negative 90 T/m tilt to the effective magnetic potential as shown by the solid red line in Fig 2 (A). Even at such a high deceleration the "front" magnetic field barrier remains at 0.25 T which is equivalent to a decelerating trap with a depth of 750mK.

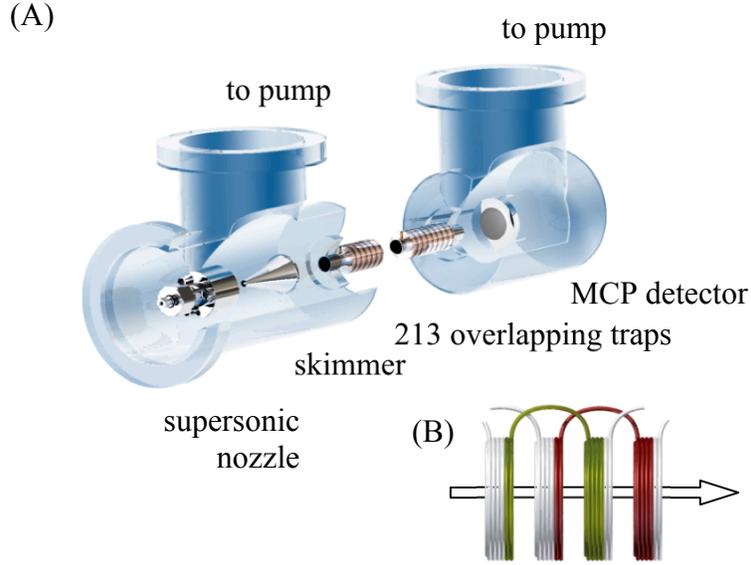

**Figure 3.** (A) Schematics of our experimental apparatus, objects are to scale, however the distances between them are not. (B) Close up view of a section of our decelerator with two overlapping quadrupole traps highlighted with color. Arrow represents the direction of beam propagation.

We now describe our experimental apparatus as shown in Fig 3. A supersonic beam of Neon is created using an Even-Lavie supersonic valve [31]. We demonstrate the operation of our decelerator using metastable neon atoms excited by a dielectric barrier discharge[32] mounted near the orifice of the valve. The long lived $^3P_2$ state of neon has two low field seeking states with mass to magnetic moment ratios of ~7 amu/$\mu_B$($m_J$=2) and ~14 amu/$\mu_B$($m_J$=1), where $\mu_B$ is the Bohr magneton. We cool the valve to 74K in order to reduce the initial beam velocity. The beam passes a 4 mm diameter skimmer mounted 15 cm from the valve. The first quadrupole trap is located 29 cm from the valve. The decelerator consists of 213 overlapping quadrupole traps extending for 114 cm. The micro channel plate (MCP) detector is located 1.4 cm from the center of the last trap and is mounted on a stage capable of a 5.2 cm translation. We extract the decelerated beam velocities and velocity distributions from time of flight measurements at different detector positions. The 1 cm bore diameter of our magnetic trap coils allows a significant simplification of our experimental setup. We place the quadrupole trap coils outside of the vacuum on a 10 mm diameter Inconel tube with a wall thickness of 0.25 mm. We chose Inconel due to its high resistivity that helps to minimize the effect of eddy currents induced by the switching magnetic fields.

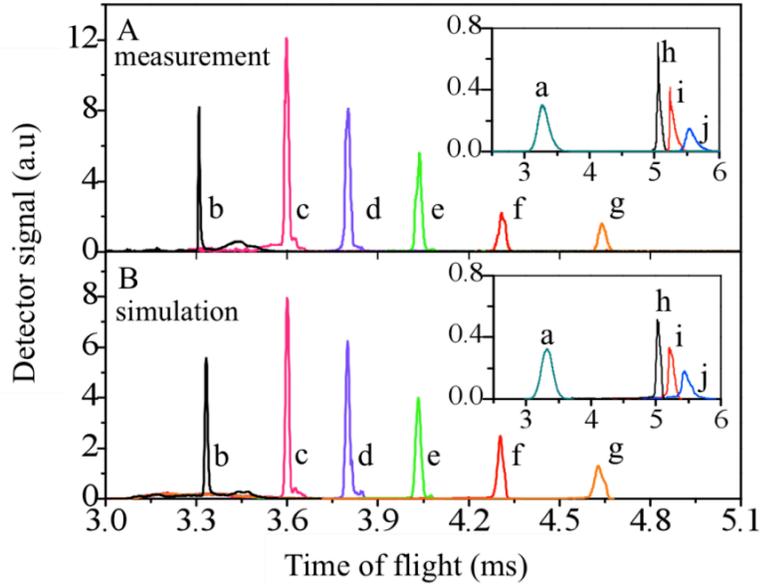

**Figure 4.** Time of flight measurements (A) and Monte-Carlo simulation results (B) of supersonic beam deceleration of metastable Neon from an initial velocity of 429.7±6.0 m/s (a–inset) to final velocities: guiding at 432.3±3.6 (b) 352.5±4.2 m/s (c), 298.6±2.2 m/s (d), 249.2±1.6m/s (e), 199.4±1.2 m/s (f), 148.4±0.8 m/s (g). Inset: lower final velocities: 97.9±1.0 m/s (h), 76.0±1.6 m/s (i), 53.8±1.1 m/s (j). The simulation was normalized to the data peak area at the final velocity of 97.9 m/s.

We measure the initial beam mean velocity to be 429.7±6.0 m/s with a standard deviation of 11.2 m/s (trace a in the inset of Fig. 4 A), corresponding to a temperature of 302mK in the moving frame of reference. In Fig. 4 we present decelerated beam intensities measured for different deceleration values. Each trace is an average of 10 measurements. As one can see the time of flight signal corresponding to the guiding mode of our decelerator (trace b Fig. 4) has a smaller area compared to the peak at 352 m/s final velocity (trace c Fig. 4). This is the direct consequence of asymmetry in our quadrupole trap. With no deceleration the lower "back" barrier height is 0.3 T (black curve Fig. 2 A). At deceleration of 28,000 m/s$^2$ the fictitious force tilts the effective potential, increasing the height of the "back" barrier to 0.4 T thus increasing the volume of our trap in the phase space. For deceleration values higher than 50,000 m/s$^2$ the "front" barrier height becomes lower than the "back" barrier height and the trap volume in the phase space starts to decrease. In our simple model the effective potential depends on deceleration only and as such we do not expect significant loss for the final velocities below 100 m/s. The 20% reduction in signal between the 100 m/s to 50 m/s velocity beams partly originates from losses that occur during the free flight to the detector that is located 1.4 cm away from the center of the last trap. Without the magnetic confinement the decelerated cloud of cold atoms expands radially to the Inconel tube walls, leading to the aforementioned loss. The free expansion loss can naturally be avoided by reducing the distance to the detector, applying a focusing coil after the last deceleration stage or by replacing the last deceleration trap with a permanent magnetic trap. From the signal area ratios we deduce that about 30% of atoms in a single quantum state ($m_J$=2) that we are able to decelerate to 350 m/s "survive" deceleration to 100 m/s as well.

It is convenient to characterize the performance of our decelerator using the acceptance parameter, the volume in the 6D phase space occupied by particles that can be slowed down to some final velocity. In order to establish the decelerator acceptance we have performed numerical Monte-Carlo simulations (no free parameters) with magnetic fields calculated using the finite element method. We calculated the longitudinal and transverse phase space distributions for two final velocities, 350 m/s (Fig. 2 C and D) and 50 m/s (Fig. 2 E and F). We start our calculations with particles uniformly distributed within a 6D block (dimensions of 60×60×60 m/s and 20×20×20 mm) in the phase space at the entrance of our decelerator. The solid lines in Fig. 2 C and E represent separatrices calculated using the one dimensional time averaged longitudinal potential including the tilt due to the fictitious force. The transverse phase space distributions are shown in Figs. 2 D and F along with separatrices constructed using the one dimensional time averaged transverse potential (Fig. 2 B). As one can see, the simple one dimensional model captures the main features of our decelerator performance. The calculated acceptance of our decelerator for slowing from 430 m/s to 350 m/s is $8 \cdot 10^6$ mm$^3$(m/s)$^3$ (close to the optimal acceptance). At a deceleration of 83,000 m/s$^2$ and a final velocity of 50 m/s the decelerator acceptance becomes $4 \cdot 10^6$ mm$^3$(m/s)$^3$. In order to operate our decelerator at low (50 m/s) velocities near its optimal acceptance we need to increase the length to 2.1 meters. Importantly, the optimal acceptance of our decelerator is very close to the emittance of our source, estimated to be about $6 \cdot 10^6$ mm$^3$ (m/s)$^3$ (at a distance of 15 cm from the valve, assuming longitudinal velocity spread of ±11 m/s and 5mm diameter skimmer 10 cm from the valve). Mode matching between the source and the moving magnetic trap can be achieved with such a configuration.

In this paper we have demonstrated a new method that is able to efficiently decelerate and stop supersonic beams of paramagnetic atoms and molecules with high acceptance. As a first step we have shown deceleration of metastable neon atoms from 429.7 m/s to 53.8 m/s in a 1.1 meter long device. We have removed 98 % of the kinetic energy.

Our method combines several important features. (i) Generality: it can be applied to decelerate both atoms and molecules that are not amenable to laser cooling, thus being very attractive to study simple reactions that are the basis of quantum chemistry. (ii) Efficiency: we avoid problems with transverse confinement by decelerating in a co-moving magnetic trap. Mode matching between the source and decelerator trap can be achieved at a price of increased decelerator length. (iii) Experimental apparatus simplicity: all of the decelerator components are placed outside of the vacuum chamber.
We expect that the combination of these features will place many interesting experiments within reach.

We acknowledge Uzi Even for many helpful discussions and sound advice. The authors would also like to thank Tal Bronshtain for the help with the experiment. We thank Christian Parthey for a careful reading of our manuscript. This research is made possible in part by the historic generosity of the Harold Perlman Family. E.N. acknowledges support from the Israel Science Foundation.**References**

1   B. C. Sawyer, B. K. Stuhl, M. Yeo, T. V. Tscherbul, M. T. Hummon, Y. Xia, J. Klos, D. Patterson, J. M. Doyle and J. Ye, *Physical Chemistry Chemical Physics*, 2011.
2   S. Willitsch, M. T. Bell, A. D. Gingell, S. R. Procter and T. P. Softley, *Physical Review Letters*, 2008, **100**, 043203.
3   S. Ospelkaus, K.-K. Ni, D. Wang, M. H. G. de Miranda, B. Neyenhuis, G. Quemener, P. S. Julienne, J. L. Bohn, D. S. Jin and J. Ye, *Science*, 2010, **327**, 853-857.
4   C. Berteloite, M. Lara, A. Bergeat, S. Le Picard, eacute, D. bastien, F. Dayou, K. M. Hickson, A. Canosa, C. Naulin, J.-M. Launay, I. R. Sims and M. Costes, *Physical Review Letters*, 2010, **105**, 203201.
5   N. Balakrishnan and A. Dalgarno, *Chemical Physics Letters*, 2001, **341**, 652-656.
6   R. V. Krems, *Physical Review Letters*, 2004, **93**, 013201.
7   I. W. M. Smith and B. R. Rowe, *Accounts of Chemical Research*, 2000, **33**, 261-268.
8   L. J. Stief, G. Marston, D. F. Nava, W. A. Payne and F. L. Nesbitt, *Chemical Physics Letters*, 1988, **147**, 570-574.
9   G. Scoles, *Atomic and Molecular Beam Methods*, Oxford University Press, USA, New York, 2000.
10  I. W. M. Smith, *Angewandte Chemie International Edition*, 2006, **45**, 2842-2861.
11  W. C. Gardiner, *Gas-Phase Combustion Chemistry*, Springer-Verlag, New York, 2000.
12  M. Gupta and D. Herschbach, *J. Phys. Chem. A*, 2001, **105**, 1626.
13  M. S. Elioff, J. J. Valentini and D. W. Chandler, *Science*, 2003, **302**, 1940-1943.
14  H. L. Bethlem, G. Berden and G. Meijer, *Physical Review Letters*, 1999, **83**, 1558.
15  N. Vanhaecke, U. Meier, M. Andrist, B. H. Meier and F. Merkt, *Physical Review A*, 2007, **75**, 031402.
16  E. Narevicius, A. Libson, C. G. Parthey, I. Chavez, J. Narevicius, U. Even and M. G. Raizen, *Physical Review Letters*, 2008, **100**, 093003.
17  R. Fulton, A. I. Bishop and P. F. Barker, *Physical Review Letters*, 2004, **93**, 243004.
18  J. M. Doyle, B. Friedrich, J. Kim and D. Patterson, *Physical Review A*, 1995, **52**, R2515.
19  B. C. Sawyer, B. K. Stuhl, B. L. Lev, J. Ye and E. R. Hudson, *The European Physical Journal D - Atomic, Molecular, Optical and Plasma Physics*, 2008, **48**, 197-209.9

20	L. Scharfenberg, H. Haak, G. Meijer and S. Y. T. van de Meerakker, *Physical Review A*, 2009, **79**, 023410.
21	E. Narevicius, C. G. Parthey, A. Libson, M. F. Riedel, U. Even and M. G. Raizen, *New Journal of Physics*, 2007, **9**, 96.
22	S. A. Meek, H. L. Bethlem, H. Conrad and G. Meijer, *Physical Review Letters*, 2008, **100**, 153003.
23	S. A. Meek, H. Conrad and G. Meijer, *Science*, 2009, **324**, 1699-1702.
24	A. Osterwalder, S. A. Meek, G. Hammer, H. Haak and G. Meijer, *Physical Review A*, 2010, **81**, 051401.
25	E. Vliegen and F. Merkt, *Journal of Physics B: Atomic, Molecular and Optical Physics*, 2005, **38**, 1623.
26	A. Trimeche, M. N. Bera, J.-P. Cromières, J. Robert and N. Vanhaecke, *Eur. Phys. J. D*, 2011.
27	W. Hänsel, J. Reichel, P. Hommelhoff and T. W. Hänsch, *Physical Review Letters*, 2001, **86**, 608.
28	M. Greiner, I. Bloch, T. W. Hänsch and T. Esslinger, *Physical Review A*, 2001, **63**, 031401.
29	G. A. Massey, D. C. Erickson and R. A. Kadlec, *Appl. Opt.*, 1975, **14**, 2712-2719.
30	L. D. Landau and E. M. Lifshitz, *Mechanics (Course of Theoretical Physics Vol. 1)* Elsevier, 1976.
31	U. Even, J. Jortner, D. Noy, N. Lavie and C. Cossart-Magos, *The Journal of Chemical Physics*, 2000, **112**, 8068-8071.
32	K. Luria, N. Lavie and U. Even, *Review of Scientific Instruments*, 2009, **80**, 104102-104104.10